\documentclass{nic-series}

\usepackage{natbib}
\usepackage{subfig}
\newcommand{\enzo}{\textit{ENZO}}

\newcommand\newblock{\hskip .11em\@plus.33em\@minus.07em}
\setcitestyle{numbers}

\begin{document} 

\title{A Song of Shocks and Dynamo: Numerical Studies of a Galaxy Cluster Merger in the HIMAG Project}

\author{Denis Wittor \inst{1,2,3} \and Paola Dom{\'{\i}}nguez-Fern{\'a}ndez \inst{3} \and Franco Vazza \inst{1,2,3} \and \\
        Marcus Br\"{u}ggen \inst{3}}

\institute{Dip. di Fisica e Astronomia, Universit\'{a} di Bologna, Via Gobetti 93/2, 40129 Bologna, Italy \\
\and
 INAF, Istituto di Radioastronomia di Bologna, Via Gobetti 101, 41029 Bologna, Italy \\
 \and
 Hamburger Sternwarte, Gojenbergsweg 112, 21029 Hamburg, Germany }

 \maketitle

\begin{abstracts}
 With {\enzo}  simulations run on the J\"{u}lich supercomputers,  we have investigated the evolution of magnetic fields in the largest cosmic structures (namely galaxy clusters and filaments connecting them) with unprecedented dynamical range. These simulations revealed the full development of the small-scale dynamo in Eulerian cosmological magneto-hydrodynamical simulations. The turbulent motions developed during the formation of  clusters are energetic enough to foster the growth of magnetic fields by several orders of magnitude, starting from weak magnetic fields up strengths of $\sim \rm \mu G$ as observed. Furthermore, shock waves are launched during cluster formation and they are able to accelerate cosmic-ray electrons, that emit in the radio wavelengths. Radio observations of this emission provide information on the local magnetic field strength. We have incorporated, for the first time, the cooling of cosmic-ray electrons when modelling this emission. In this contribution, we present our advances in modelling these physical processes. Here, we mostly focus on the most interesting object in our sample of galaxy clusters, which shows the  complexity of magnetic fields  and the potential of existing and future multi-wavelengths observations in the study of the weakly collisional plasma on  $\sim$ Megaparsecs scales. 
\end{abstracts}

\section{Introduction}
 Galaxy clusters are the largest plasma laboratories in the Universe. They contain hundreds of galaxies that sit at the nodes of the cosmic web, which consists primarily of Dark Matter $(\sim 83 \ \%)$, as well as a hot, ionized, weakly collisional plasma called the intracluster medium (ICM, $\sim 26 \ \%$). Note that stars only make up about a few percent of the total mass of galaxy clusters. \\
 While the star light of the cluster is observable in the optical light, the ICM is not and one has to use X-ray and radio observations to shed light on the state and properties of this plasma. The ICM hosts a variety of thermal and non-thermal phenomena that are driven during the evolution of the cluster by the accretion of matter and via the merging with other clusters. These mergers drive both shock waves and turbulence in the ICM \citep[e.g.][]{ry03,mi14}. Moreover, radio observations reveal the presence of diffuse radio emission on large-scales, giving evidence for the presence of large-scale magnetic fields and relativistic particles. Shock waves are expected to accelerate cosmic-ray electrons via diffusive shock acceleration and to produce radio relics, while turbulence is assumed to produce radio halos via turbulent re-acceleration of cosmic-ray electrons. Radio relics are large and elongated sources that are found on the cluster periphery. Moreover, relics have a high degree of polarisation. On the other hand, radio halos are diffuse sources located at the cluster center and they do not show any sign of polarised emission \citep[e.g.][for a recent review]{2019SSRv..215...16V}. \\
 Many questions still surround the ICM. While both radio relics and halos are produced by cosmic-ray electrons, there has not been a signal, in form of $\gamma$-rays, of cosmic-ray protons that should be accelerated by the same mechanisms \citep[e.g.][]{wi17}. Moreover, it is still debated if the origin of the large-scale magnetic fields is primordial or astrophysical \citep[e.g.][]{va17cqg}. Since the characteristic time scales of the systems are large, $\sim \mathrm{Gyr}$, it is impossible to study their evolution using observations. Hence, realistic numerical simulations are required to connect theory and observations in a quantitative way.\\
 Simulating the ICM poses some technical and computational challenges for which different numerical algorithms have been designed. Beside the needs of properly resolving the scales at which Dark Matter and baryonic matter form clusters, the study of non-thermal phenomena in the ICM and their related radio emission makes it mandatory to resolve in time and space the small-scale dynamo amplification of magnetic fields, triggered by turbulent motions \citep[e.g.][]{review_dynamo}. Hence, the physical scales that need to be considered span several orders of magnitude, i.e. going from the Giga-parsec to the parsec\footnote{One parsec is $3.1 \times 10^{16} $ meter.} scale. When resolving these small scales, additional physical process related to galaxy formation (such as feedback from active galactic nuclei or radiative processes) must be included. Hence, these computational challenges can only be tackled by using high-performance computing.  \\
 In this contribution, we present our recent contributions to the study of galaxy clusters and magnetic fields within them using high-resolution simulations and supercomputing at the J\"{u}lich FZ.  First, will present our numerical methods and benchmark tests performed on JUWELS in Sec. \ref{sec_forces}. In  Sec. \ref{sec_design}, we present our cosmological simulation in greater detail and provide an overview over our scientific highlights. We give our concluding remarks in Sec. \ref{sec_conclusion}.
\section{Numerical methods \& Hardware considerations} \label{sec_forces}
\subsection{Cosmological Magneto-Hydrodynamics with ENZO}
 Several numerical codes have been developed to simulate the cosmic web and its components. In this work, we used the cosmological grid code {\enzo} \citep[][]{enzo13}. {\enzo} is an adaptive mesh refinement (AMR) code that solves the equations of magneto-hydrodynamics (MHD) in the cosmological framework. It is being developed as a collaborative effort of scientists at many universities and national laboratories. {\enzo} uses a particle-mesh N-body method (PM) to follow the dynamics of the collision-less Dark Matter component, and an adaptive mesh method for ideal fluid dynamics. The Dark Matter component is coupled to the baryonic matter (gas) via gravitational forces, calculated from the total mass distribution (Dark Matter + gas) by solving the Poisson equation with an FFT based approach. In its basic version, the gas component is described as a perfect fluid ($\gamma = 5/3$) and its dynamics are calculated by solving conservation equations of mass, energy and momentum over a computational mesh, using a Eulerian solver based on the Piecewise Parabolic Method (PPM) or the lower order Piecewise Linear Method (PLM). These schemes are higher-order extensions of Godunov's shock capturing method.\\
 {\enzo} is parallelized by domain decomposition into rectangular sub-grids, including the top/root grid (which is the only level in a non-AMR run). Message passing paradigm is adopted and implemented by means of the MPI library (see http://www.mpi-forum.org/), while I/O makes use of the HDF5 data format (see http://www.hdfgroup.org/HDF5/).\\
 The MHD extension used in our runs is based on the conservative Dedner formulation of MHD equations \citep[][]{ded02}, which uses hyperbolic divergence cleaning to preserve the divergence of the magnetic field as small as possible. The MHD solver adopts the PLM reconstruction, fluxes at cell interfaces are calculated using the local Lax-Friedrichs Riemann solver (LLF) and time integration is performed using the total variation diminishing (TVD) second-order Runge-Kutta (RK) scheme. The resulting solver is somewhat more diffusive than the PPM approach, but allows a more efficient treatment of the electromagnetic terms.
\begin{figure}
 \centering
  \includegraphics[width = 0.45\textwidth,height=0.42\textwidth]{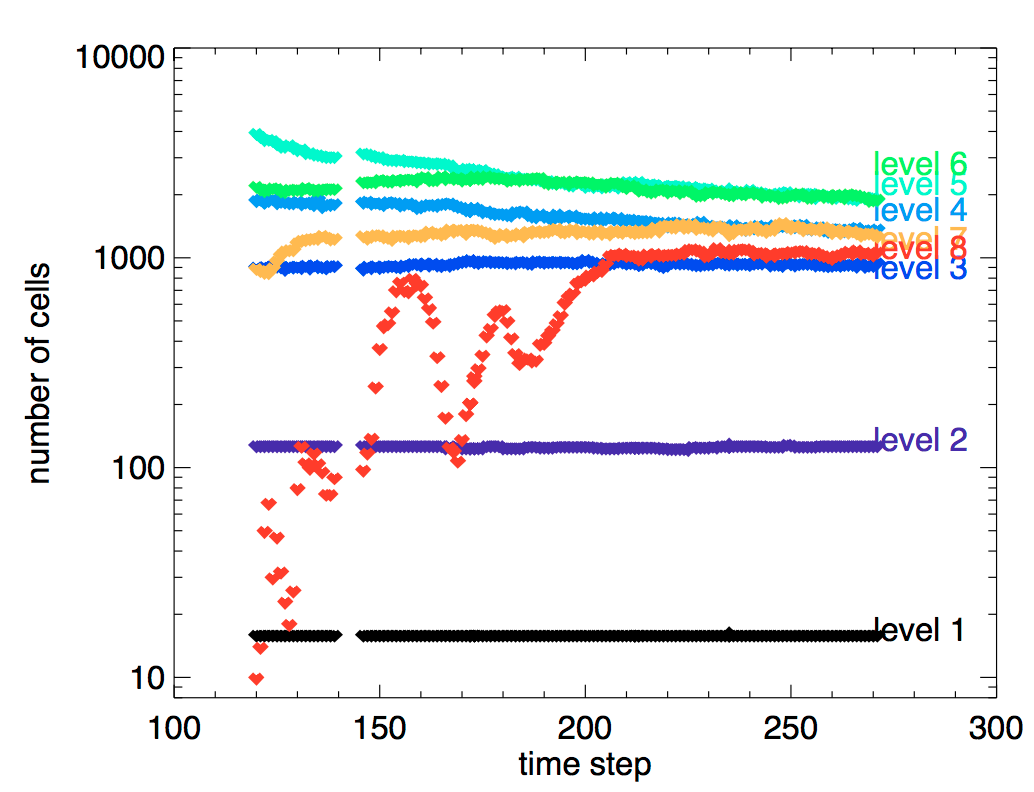}
   \includegraphics[width = 0.45\textwidth,height=0.40\textwidth]{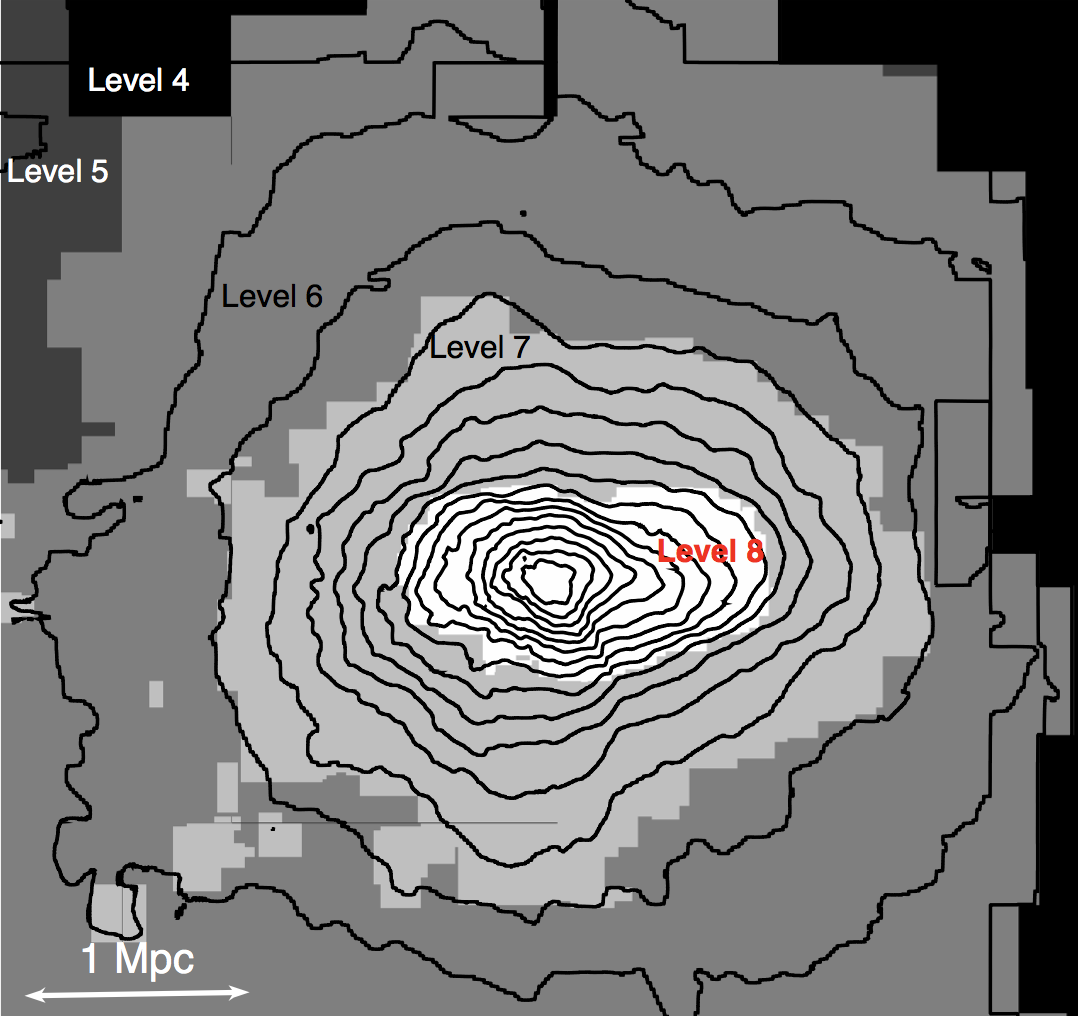}
    \includegraphics[width = 0.45\textwidth,height=0.38\textwidth]{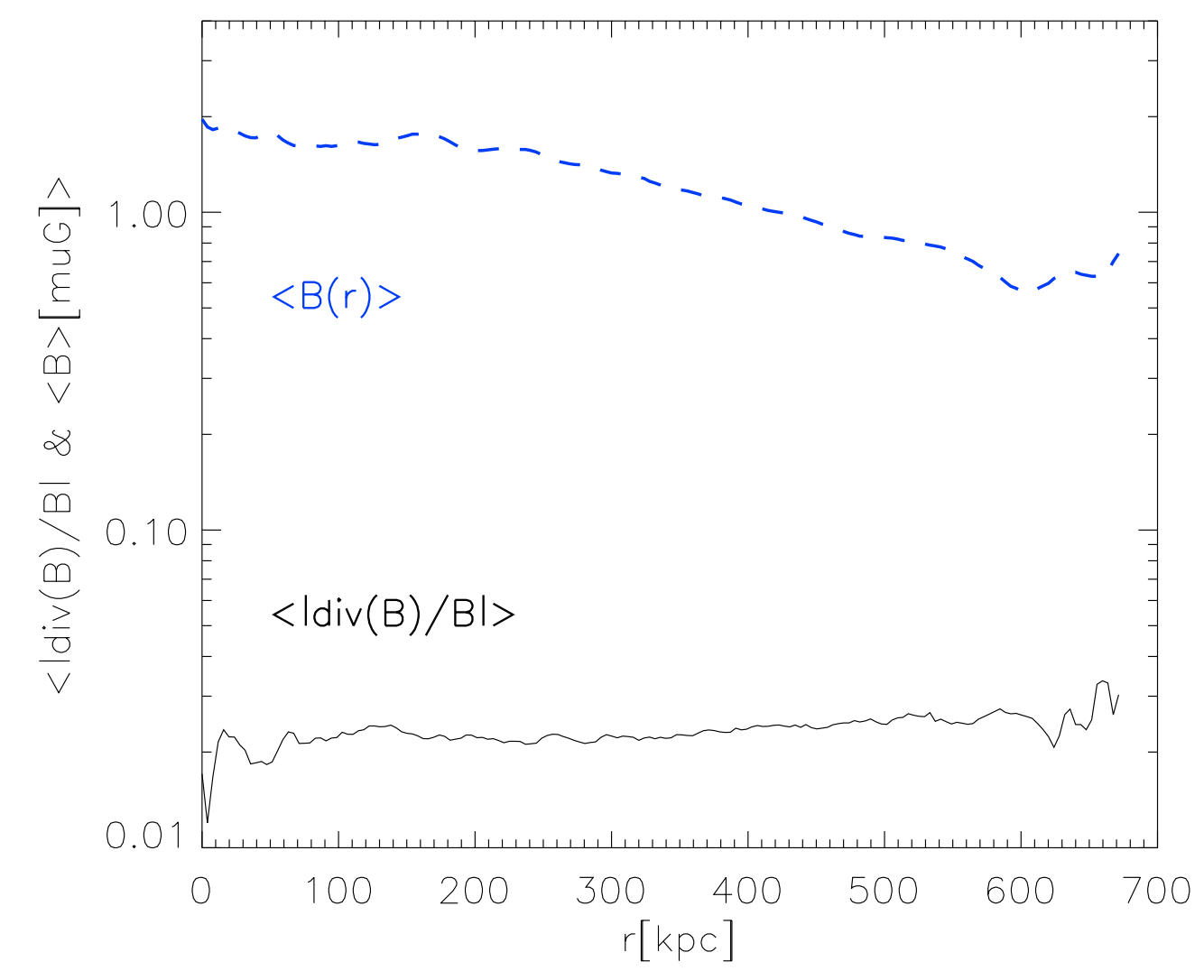}
  \includegraphics[width = 0.45\textwidth,height=0.40\textwidth]{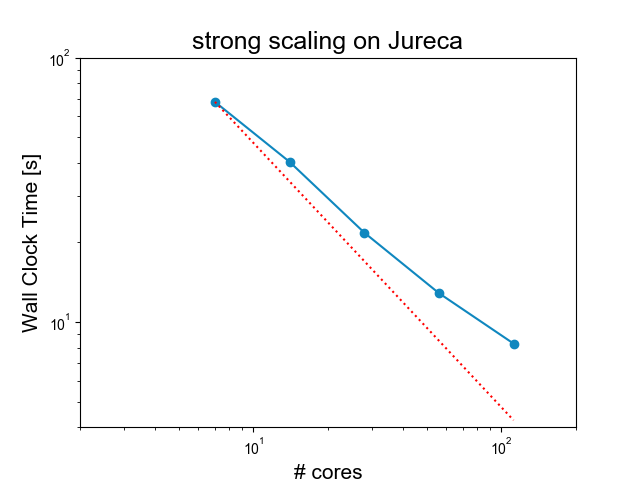}
 \caption{A few computing/numerical tests on our AMR strategy. Top left: progression in the number of cells at different AMR levels produced by a typical 8-levels {\enzo} cluster simulation on JUWELS. Top right: thin slice giving the distribution of AMR levels through the center of a simulated cluster. Bottom left: radial profile of magnetic field and residual divergence of magnetic field in a simulated cluster at $z=0$. Bottom right: strong scaling test on JURECA.} 
 \label{fig1}
\end{figure}
\subsubsection{Adaptive Mesh Refinement Strategy}
 The mesh refinement approach in {\enzo} adopts an adaptive hierarchy of grids at varying levels of resolution. \enzo's implementation of structured AMR poses no fundamental restrictions on the number of grids at a given refinement level or on the number of refinement levels. Moreover, it also allows arbitrary integer ratios of parent and child grid resolution. The patches are organized through a Hierarchical Tree structure, responsible for the management of adding or removing patches, specifying the neighbouring boxes at the same refinement level, the parent of each box, etc. In {\enzo}, no parallelisation is implemented on the Hierarchical Tree and the hierarchy data structure has to be replicated on each processor. This often yields a a large memory usage overhead for our production runs.\\
 In order to cope with the memory bounds on computing nodes on JURECA and JUWELS, we imposed a maximum AMR level of 8, considering a fixed refinement ratio of 2, which gave a $2^8=256$ refinement of the initial mesh resolution, i.e. from $\Delta x_0=1014 \rm ~kpc$ to $\Delta x_8=3.9 \rm ~kpc$. It shall be noticed that obtaining the highest possible spatial resolution when dealing with small-scale dynamo of magnetic fields is mandatory, as a coarse spatial resolution limits the effective Reynolds number of the simulated flows, and hence can severely hamper the development of the small-scale dynamo \citep[e.g. see discussion in ][]{review_dynamo}.  \\
 In order to maximize the dynamical range for each simulated object, in HIMAG/2 we started from two levels of static uniform grids with $256^3$ cells each and using $256^3$ particles each to sample the Dark Matter distribution, with a mass resolution per particle of $m_{\rm DM}=1.3 \cdot 10^{10} M_{\odot}$ at the highest level. Then, we further refined the innermost  $\sim$ (25 Mpc)$^3$ volume, where each cluster forms, with additional 7 AMR levels (refinement $=2^7$).  The refinement was initiated wherever the gas density was $\geq 1\%$ higher than its surroundings. This ensures that most of the central volume of each simulated clusters (where the small-scale dynamo is expected to develop while gas and dark matter continue to be assembled via mergers and accretions) is resolved up to the maximum possible resolution, in a rather uniform way. Figure \ref{fig1} shows the increase in the number of levels as a function of time in a typical cluster simulation in HIMAG/2, as well as the map of AMR levels in a cluster at the end of the simulation: the innermost $\approx 2^3 \rm ~Mpc^3$ is typically resolved down to $\Delta x_8=3.9$ kpc, while the entire cluster volume is refined from $\Delta x_7=7.8$ kpc to $\Delta x_8=15.6$ kpc. With this approach, our AMR strategy was able to limit the spurious generation of the divergence of magnetic fields to a few percent in the entire volume of simulated galaxy clusters, as shown in the bottom left panel of Fig.\ref{fig1}.
\subsubsection{Optimisation for J\"{u}lich Supercomputers} \label{sec_benchmark} 
 JURECA and JUWELS are well-suited for this project, given that the AMR implementation in {\enzo} requires a relatively small number of cores but a large memory per node. The bottom right panel of Fig.~\ref{fig1} shows an example of a strong scalability test for a $256^3$ root grid, with AMR triggered by ovedensity refinement for 3 AMR levels in the innermost $64^3$ region, using from 7 to 112 cores.  The behaviour of the code after 200 timesteps  is very stable, and the average wallclock time per timestep is a factor $\sim 2$ above the theoretical scaling for the 112 cores case.\\
 In full production, we used up to $\sim 680$ cores on 64 nodes to cope with the memory bounds posed by our runs, which can require up to  $\sim 1.8$ Tb of memory towards the end of the simulation, due to the large number of cells generated by the AMR scheme and by the data stored in the hierarchy tree. Each cluster of our sample was simulated independently, using $\sim 30,000-50,000$ core hours on JURECA (HIMAG) and JUWELS (HIMAG2). 
\begin{figure}
 \includegraphics[width = 0.49\textwidth]{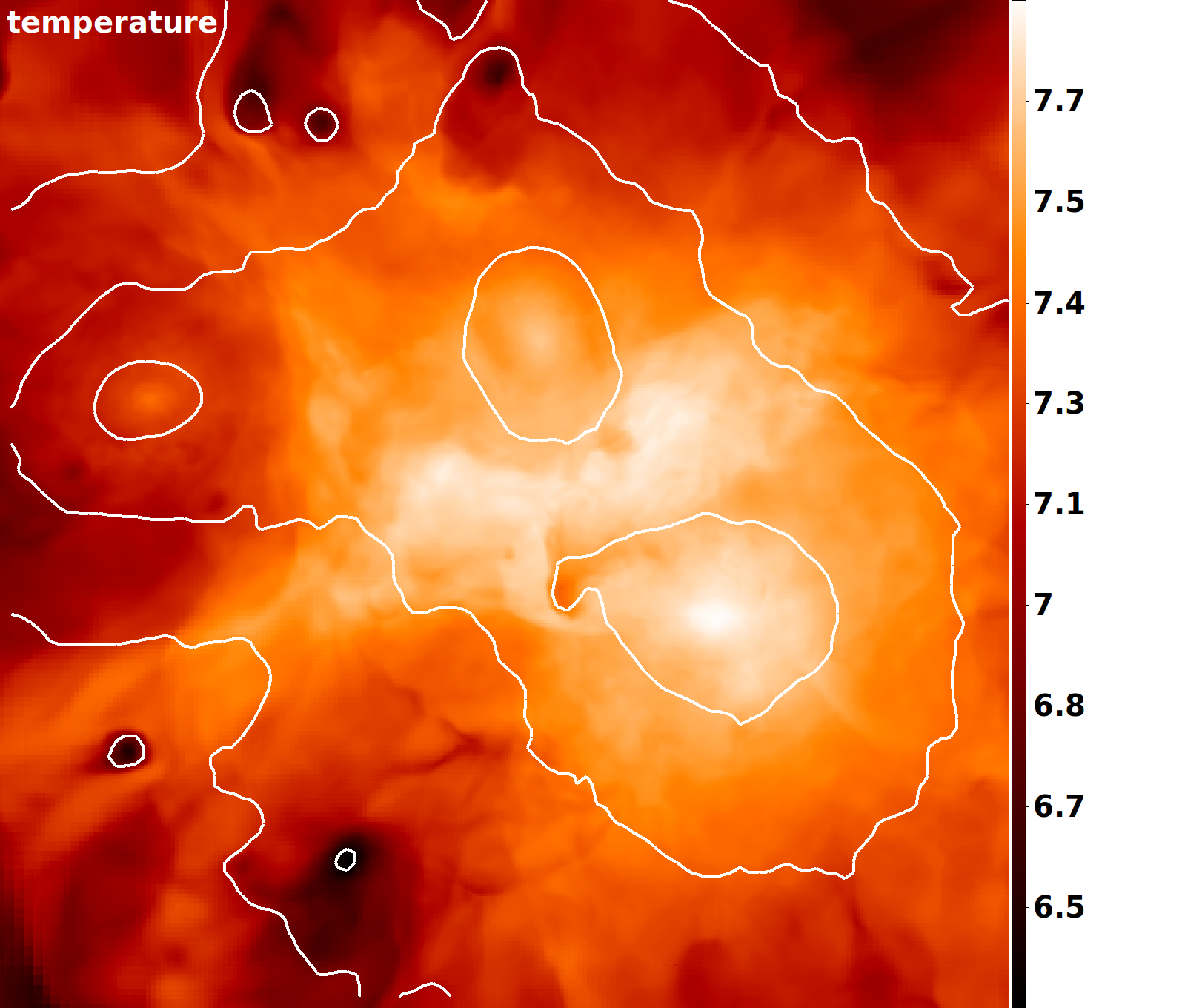} 
 \includegraphics[width = 0.49\textwidth]{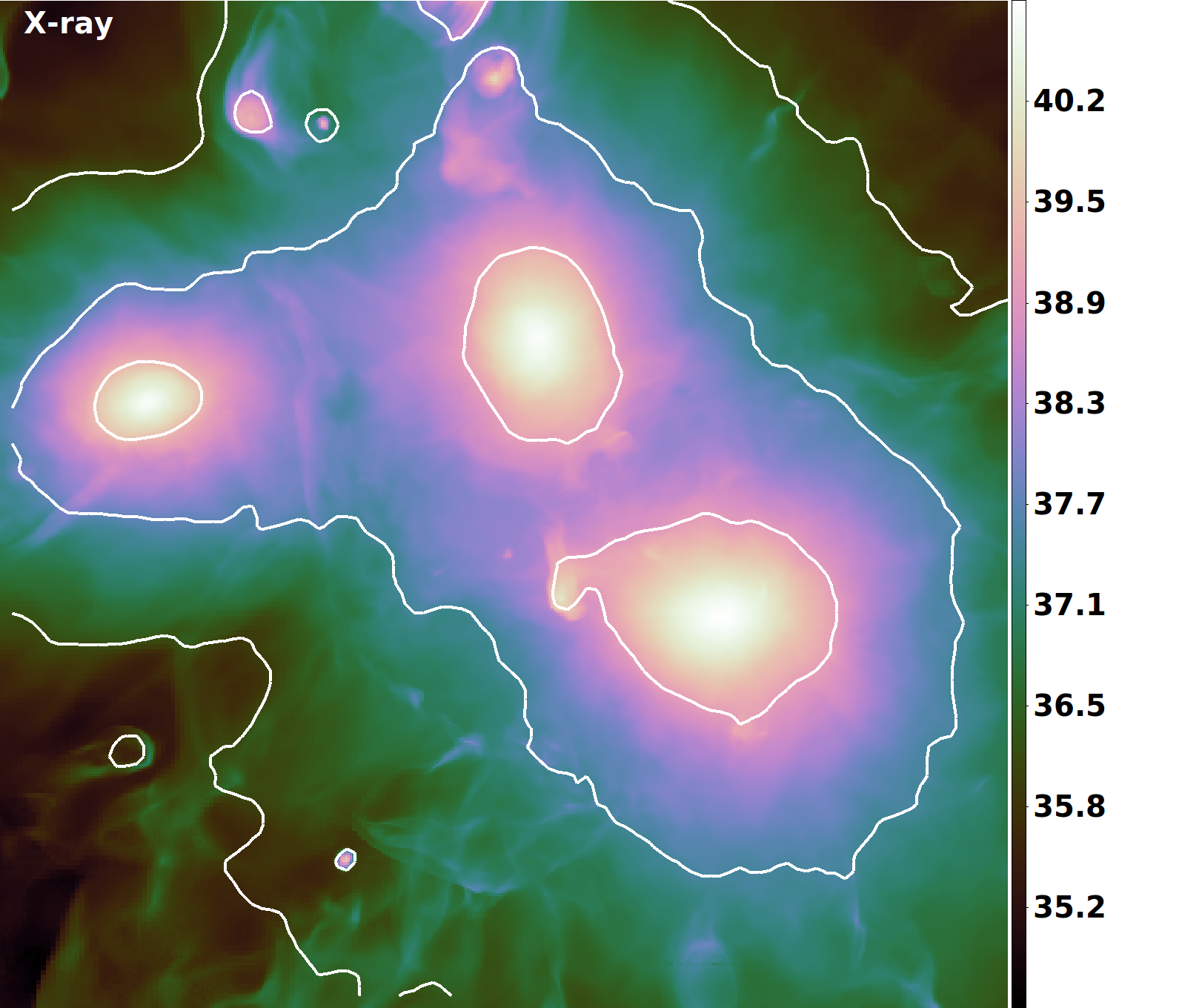}
 \caption{Projected gas temperature (left), in K, and X-ray luminosity (right), in erg/s, of the simulated galaxy cluster E5A. Both maps are given in log-scaling and they are overlayed with the baryonic density contours.}
\label{fig:e5A}
\end{figure}

\section{Scientific results}\label{sec_design}
 Using the above strategies, we simulated a set of galaxy clusters within the HIMAG/2 projects. The main results were presented in several papers: \cite{va18mhd}, \cite{dom19}, \cite{hack18}, \cite{va19},  \cite{2019Sci...364..981G}, \cite{2019MNRAS.tmp.2080S} and \cite{wi19}. Here, we want to highlight some of our results by focusing on one peculiar cluster in our sample (identified as "E5A") which displays a number of remarkable properties\footnote{See https://vimeo.com/266941122 for a movie showing the evolution of gas density and magnetic fields for this system.}. Fig.~\ref{fig:e5A} shows the projected gas temperature and the projected X-ray luminosity of E5A at the epoch of $z \approx 0.1$. \\
 For the simulation of E5A, as well as for all other simulations, we assumed the same $\Lambda$CDM cosmological model, with $h = 0.72$, $\Omega_{\mathrm{M}} = 0.258$, $\Omega_{\mathrm{b}}=0.0441$ and $\Omega_{\Lambda} = 0.742$ \citep[see e.g.][]{dom19}.  A public repository of cluster snapshots from this catalog is available at this URL (https://cosmosimfrazza.myfreesites.net/amr\_clusters).
\subsection{The evolution of magnetic fields in an interesting cluster pair}
 In \cite{va18mhd}, we  have presented evidence for {\it resolved} dynamo growth of intracluster magnetic fields in a Coma-like galaxy clusters from our sample. The unprecedented dynamical range achieved in the innermost cluster regions provided evidence of a small-scale dynamo and local amplification of magnetic fields up to values similar to what is found in observations ($\sim \mu G$), approaching energy equipartition with the kinetic energy flow on $\leq$ 100 kpc scales. We could constrain an overall efficiency of order $\sim 4 \ \%$ in the transfer between turbulent kinetic energy (in the solenoidal component) and the magnetic energy field.  \\
 {\bf Furthermore, the topology of the magnetic fields seem to be consistent with observations of the Coma cluster \citep[e.g.][and references therein]{bo13}}. In particular, the Faraday Rotation of background polarised sources is in good statistic agreement with the observations of real sources in the background of Coma, even if our simulation produce significant non-Gaussian tails in the distribution of magnetic fields components, which stems from the superposition of different amplification patches which mix in the ICM. This result has an important implication since a Gaussian distribution of magnetic field components is often assumed in the interpretation of Faraday Rotation \citep[e.g][]{bo13}.
\begin{figure}
    \centering
    \includegraphics[width=0.49\textwidth]{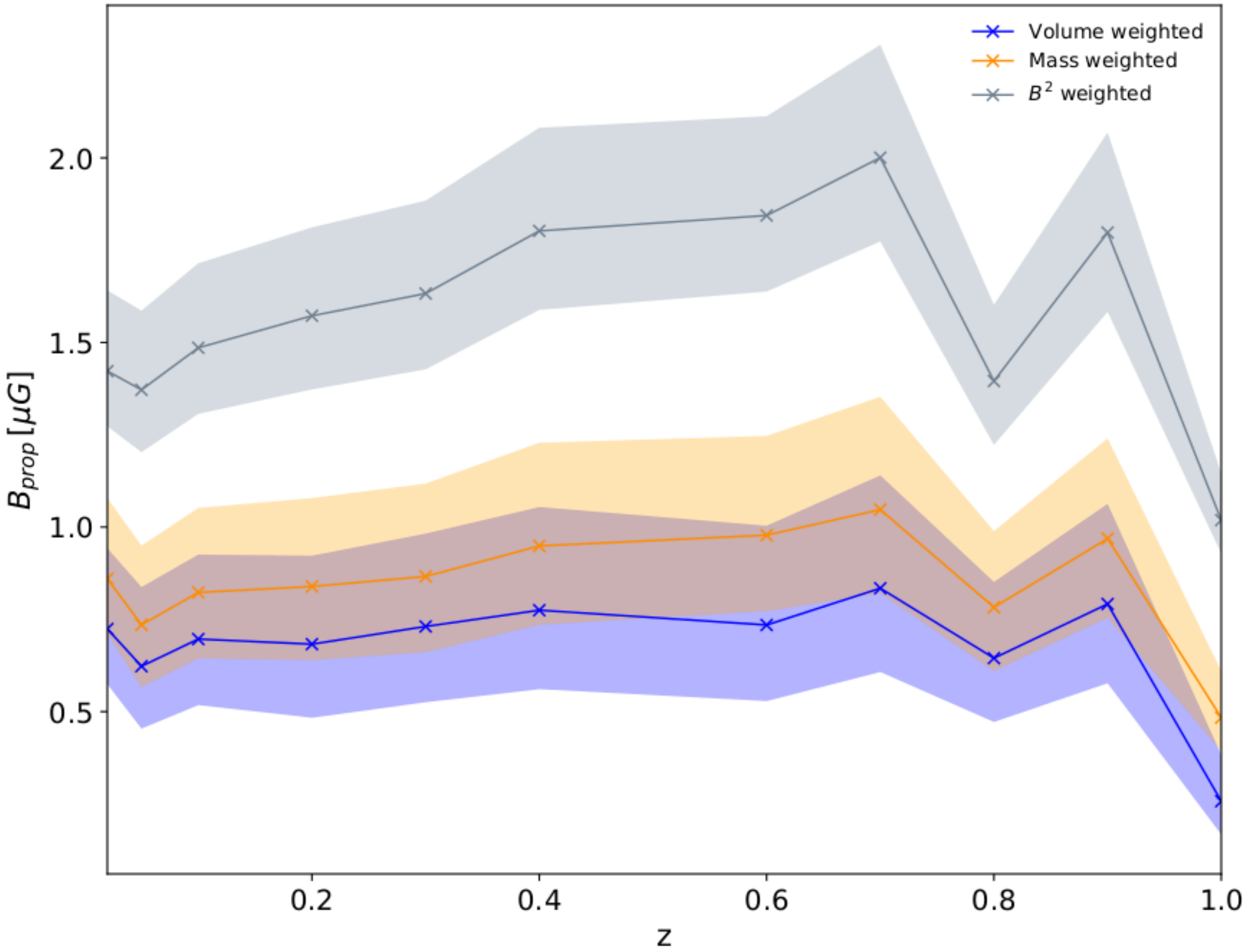}
    \includegraphics[width=0.49\textwidth]{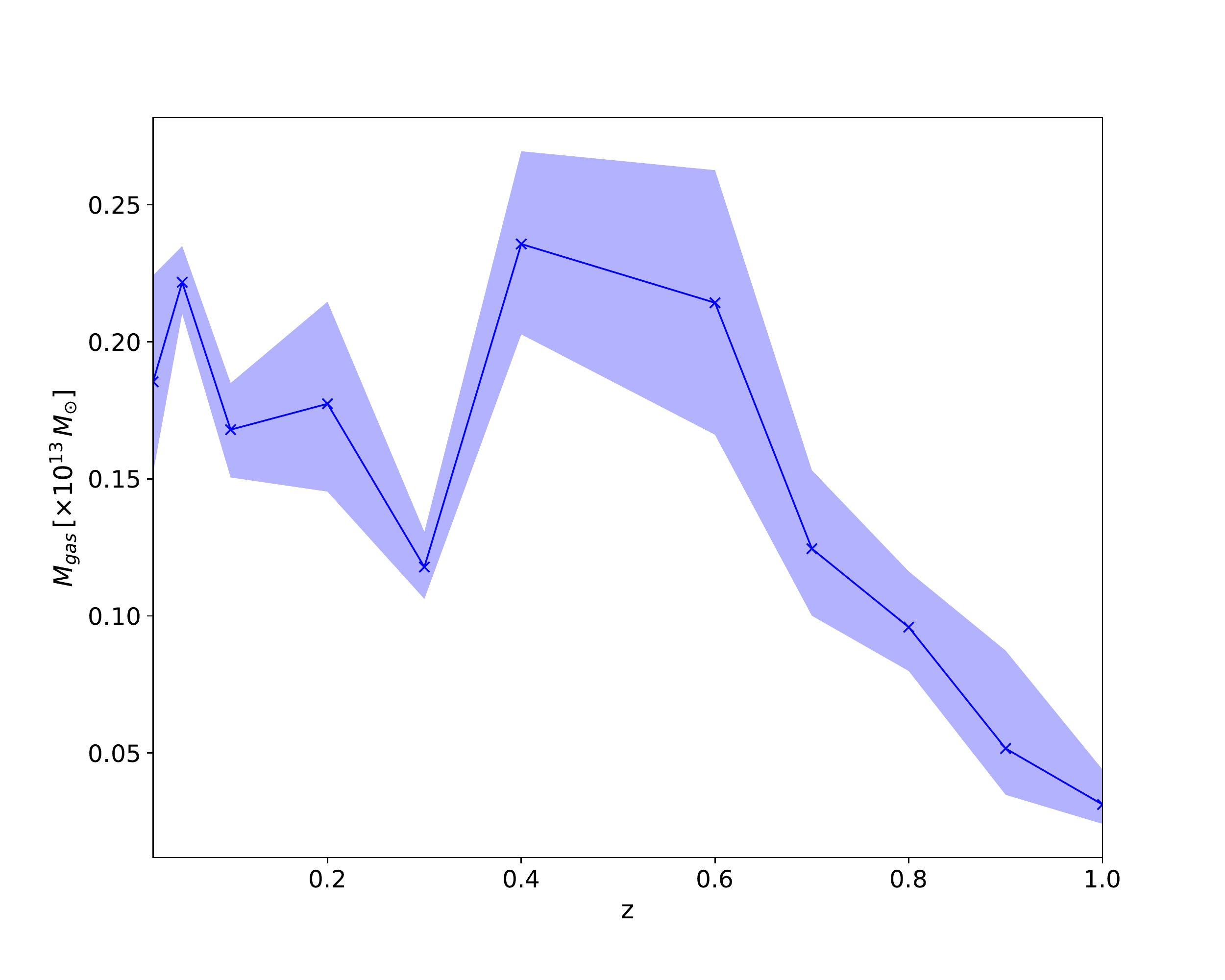}  \\
    \includegraphics[width = 0.49\textwidth]{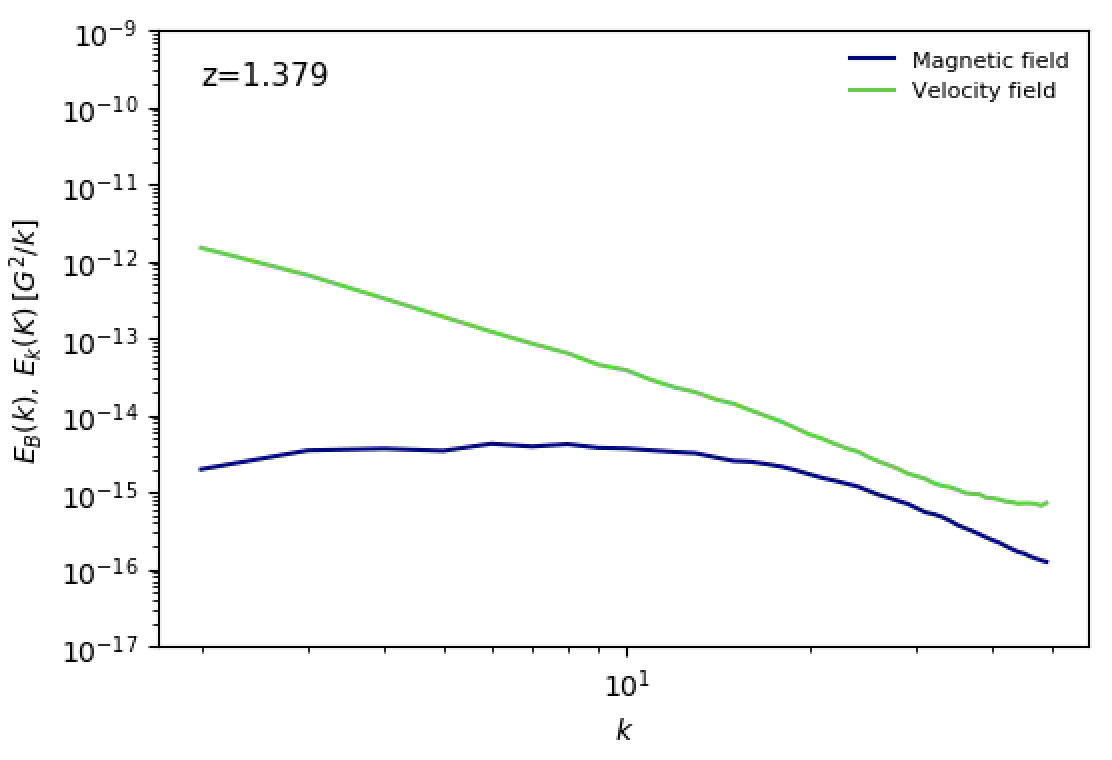}
    \includegraphics[width = 0.49\textwidth]{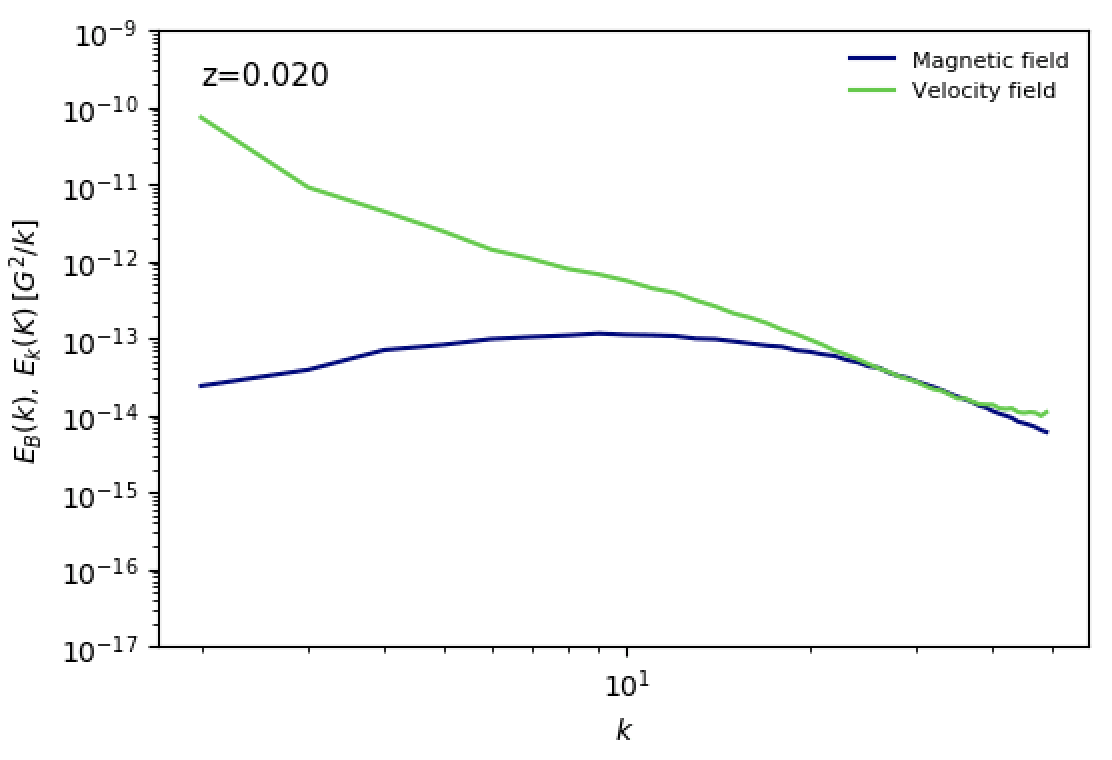}
    \caption{Top panels: evolution of the physical magnetic field strength (left) and gas mass (right) of E5A with redshift. The magnetic field is shown with a volume, mass and $B^2$ weight. The shadowed areas in both panels correspond to scatter of values within a radius of $1 ~\rm Mpc$ at each redshift. Bottom panels: E5A magnetic (purple) and kinetic (green) spectra evolution at different redshifts.} 
     \label{fig:Spec}
 \end{figure}\\
 In \cite{dom19},  we generalized this analysis to a larger set of clusters and snapshots, produced in HIMAG and HIMAG2. There we studied how the spectral properties of magnetic fields are affected by mergers, and we could relate the measured magnetic energy spectra to the dynamical evolution of the ICM.  Our results show that the magnetic growth rate is larger for merging systems and that relaxed systems generally have a larger outer scale. Since this is the scale that characterizes the magnetic spectrum, larger values of the outer scale suggest that a dynamo has acted for a longer time in such systems. In \cite{dom19}, we also studied in more detail the evolution of E5A. As an example, we show in the top panels of Fig. \ref{fig:Spec} how the physical magnetic field and the gas mass of the cluster are changing with redshift (analysis made in a $\rm Mpc^3$ cube moving with the cluster center of mass). The magnetic energy of E5A grows by a factor of $\sim 40-50$ in a time-span of $\sim 9$ Gyr and equipartition between kinetic and magnetic energy occurs on a range of scales ($< 160 \rm ~kpc$ at all epochs), depending on the turbulence state of the system. Several minor-merger and major-merger events affected the magnetic properties of E5A cluster and shaped its magnetic spectrum in a complicated manner (see bottom panels of Fig. \ref{fig:Spec}).  During mergers, the peak in the magnetic spectra shifts to {\it smaller spatial scales} and thus the magnetic spectrum broadens. A major-merger event affects the magnetic amplification of the cluster by delaying it for $\lesssim$ 1 Gyr, in contrast, continuous minor mergers seem to promote the steady growth of the magnetic field.
\subsection{Magnetic fields illuminated by cosmic shocks }
 In \cite{wi19} and \cite{2019MNRAS.tmp.2080S}, we used two clusters of the HIMAG/2 sample to study the properties of radio relics. To compute their radio relic emission in cosmological simulations, one has to follow a simple recipe: First, one has to find shock waves in the simulation. Therefore, we used \textit{Velocity-Jump} method that searches for jumps in the three-dimensional velocity field to detect shock waves \citep[][]{va19}. Second, one has to assign the radio emission to the detected shock. Cosmic-ray electrons are expected to have short live times ($\le 10 \ \mathrm{Myr}$) and, hence, they loose most of their energy within $100 \ \mathrm{kpc}$ of their injection site. Moreover, their energy budget seems to be dynamically unimportant. Hence, one can assume a quasi-stationary balance between the shock acceleration and radiative losses and, compute the cosmic-ray electron energy budget and associated radio power using the information of a single timestep \citep{hb07}. \\
 With the increasing resolution of cosmological simulations, the assumption of a quasi-stationary balance becomes invalid as the resolution drops below the electron cooling length. Hence, one has to properly resolve the downstream profile of the radio emission. In \cite{wi19}, we have developed an algorithm that takes the cooling of electrons into account and attaches a radio emission profile to each shocked cell. In addition to it, we have implemented an algorithm, following \cite{burn66}, that computes the polarised emission of radio relics. For details on the implementation we point to \cite{wi19}. In the left panel of Fig. \ref{fig_radio}, we give an example of the polarisation fraction of E5A. Using the new model, we have produced the most detailed simulation to date of radio polarisation in cluster shocks in \cite{wi19}  and \cite{2019MNRAS.tmp.2080S}.  \\
 In \cite{wi19}, we found that {\bf the observed polarised emission from radio relics should strongly depend on the environment}, and that the direction of polarisation and its variation across the relic depend on the properties of the upstream gas. Laminar gas flows in the upstream result in a parallel alignment of polarisation vectors, while disturbances in the upstream will cause a random orientation. These observations might reflect the local correlation length of the magnetic field. Furthermore in \cite{2019MNRAS.tmp.2080S}, we combined VLA observation of the western relic in RXCJ1314.4-2515 and cosmological simulations of a similar relic to study the history of the radio relic. We found that the relic power, measured at 3 GHz, can only be explained by the re-acceleration of an fossil population of cosmic-ray electrons.
\subsection{Intracluster bridges}
\begin{figure}
 \centering
 \includegraphics[width = 0.49 \textwidth]{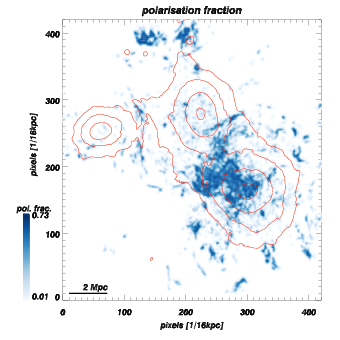}
 \includegraphics[width = 0.49\textwidth]{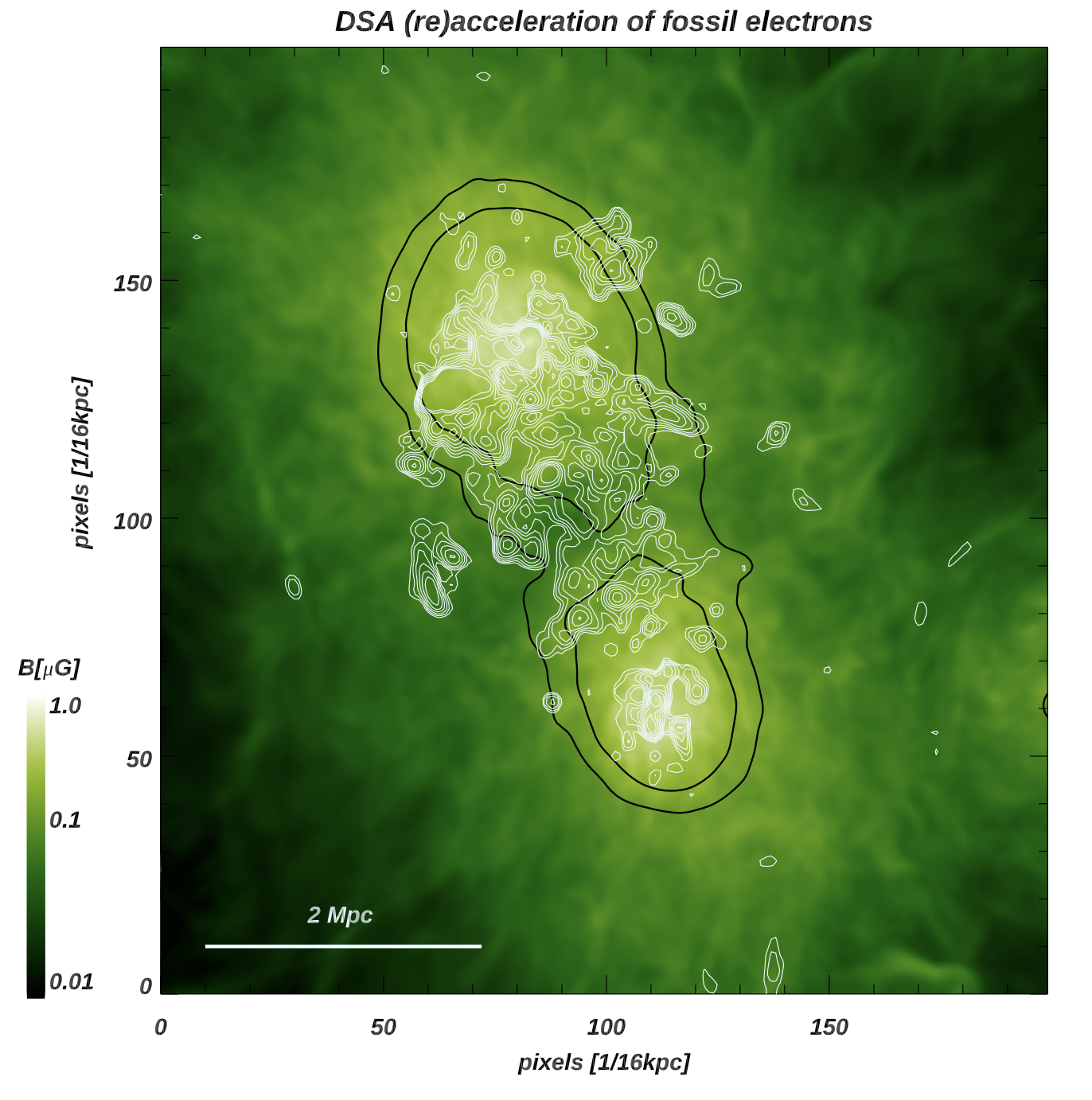}
 \caption{Left: Density contours and polarisation fraction (colors) of E5A. Right: Projected magnetic fields (color), radio emission at 200 MHz (white contours) and projected Sunyaev-Zeldovich decrement (black contours) for the same object. } 
 \label{fig_radio}
\end{figure}

 Massive binary mergers as the one in E5A are rare and powerful events, in which a large amount of kinetic energy is concentrated between the two main clusters,  and  gas matter gets strongly compressed in the region connecting them.  This collapse is rather fast ($\leq \rm Gyr$), leading to transonic turbulence and weak shocks \citep[][]{va19}. This process leads to the compression and {\bf "boosting" of the X-ray emission of the Warm Hot Intergalactic Medium  (WHIM)} in the interacting regions, which is otherwise invisible. \\
 We  produced synthetic radio and X-ray observations of E5A, tailored to reproduce existing (LOFAR, MWA and XMM) or future instruments (SKA-LOW and SKA-MID,  ATHENA and eROSITA). Our simulations show that with sufficiently long ($\sim 10^2 \ \rm{rm}~ ks$) exposures in the soft ($0.3-1.2$ keV) X-ray band on candidates similar to E5A it should be possible to {\bf measure important thermodynamic properties of intracluser bridges, providing complementary constraints to radio data.} For example, with X-IFU it should be possible to  derive the Mach number of radio shocks entirely from spectroscopically-derived information of the local gas velocity dispersion and of the local sound speed, in a temperature regime which is difficult to find in galaxy clusters \citep[][]{va19}.\\ 
 As a follow up, in \cite{2019Sci...364..981G} we  mock-observed E5A with a LOFAR-HBA configuration, and compared it to real observations of the A399-A401 system of galaxy clusters. \cite{2019Sci...364..981G} reported the LOFAR-HBA detection of spectacular ridge of radio emission extending several Megaparescs between A399 and A401. Using E5A as a testbed for particle acceleration models, we studied under which circumstances the intergalactic magnetic fields between the two clusters could become bright enough at radio wavelengths to match the real LOFAR observation (see right panel of Fig. \ref{fig_radio}). We reported that standard Diffusive Shock Acceleration alone cannot provide enough acceleration of cosmic-ray electrons here to explain the observed radio emission, and thus that some more volume filling (and/or) efficient acceleration mechanism should operate in the intergalactic medium. This implies that systems like E5A, although rare in the Universe, can be more radio luminous than expected, while remain challenging to be detected in other wavelengths.
\newpage
\section{Concluding Remarks}\label{sec_conclusion}
Achieving a large dynamical range in the numerical description of evolving magnetic fields in large-scale structures is crucial to allow the quantitative modelling of complex radio data. Thanks to the allotted HIMAG/2 project on JURECA and JUWELS, our group produced the most resolved MHD simulation to date of magnetic fields in galaxy clusters with {\enzo} \citep[][]{enzo13}, focusing with unprecedented detail on the evolution of intracluser magnetic fields under the action of the small-scale turbulent dynamo \citep[][]{va18mhd,dom19}.  \\
 Among the investigated systems, the simulated merger in E5A has a peculiar evolutionary track, owing to its complex cycle of shock waves, turbulence injection and dynamo amplification episodes. As discussed in this contribution, E5A has thus highlighted a number of very interesting and little explored features of magnetic fields in cosmology:  from the sequence of complex amplification cycles following mass accretions \citep[][]{dom19} to the powering of polarised radio emitting shocks \citep[following][but not analysing E5A]{wi19}, and from the association of detectable radio and X-ray features in intracluser bridges \citep[][]{va19} to the hints for more efficient particle acceleration mechanisms in such regions \citep[][]{2019Sci...364..981G}. \\
 Despite the improvements in numerical study of galaxy clusters presented in this work, there are still several physical processes, such as AGN feedback and galaxy formation, that have to be incorporated, making the simulations more complex and computational challenging. Thanks to the availability of high-performance computers, such as JUWELS, it will be possible to overcome these challenges and to make new exciting scientific discoveries in the future.
\section*{Acknowledgments}
  The cosmological simulations described in this work were performed using the {\enzo} code (http://enzo-project.org).  The authors acknowledge the usage of computing time through the John von Neumann Institute for Computing (NIC) on the GCS Supercomputer JUWELS at J\"ulich Supercomputing Centre (JSC), under projects no. 11823, 10755 and 9016 as well as hhh42,  hhh44 and stressicm.   D. W.,  F.V. and P. D. F. acknowledge financial support from the European Union's Horizon 2020 program under the ERC Starting Grant "MAGCOW", no. 714196. We also acknowledge the usage of online storage tools kindly provided by the Inaf Astronomica Archive (IA2) initiative (http://www.ia2.inaf.it). 
\footnotesize{}
\bibliographystyle{nic}
\bibliography{franco.bib}

\end{document}